\documentclass[10pt]{article}
\usepackage{amsmath}   
\usepackage{amsfonts}  
\begin{document}
\vspace*{2.5cm}
{\Large\bf Absolute Spacetime:\par
The Twentieth  Century
Ether\par }\bigskip\bigskip

\vspace*{0.8cm}
{\bf Carl H. Brans\footnote{Loyola University, New
Orleans, LA 70118, email:brans@loyno.edu}}
\vspace*{0.5cm}
\hspace*{1.5cm}\begin{minipage}[t]{9.5cm}
\footnotesize{\it Received }

\hrulefill

  All gauge theories need ``something fixed'' even as ``something
changes.''
Underlying the implementation of these ideas all major physical
theories make indispensable use of an elaborately designed
spacetime model as the ``something fixed,'' i.e., absolute.
This model must provide at least the following sequence of
structures: point set, topological space, smooth manifold,
geometric manifold, base for various bundles. The ``fine
structure'' of spacetime inherent in this sequence is
of course empirically unobservable directly, certainly when
quantum mechanics is taken into account. This issue is at the
basis of the difficulties in quantizing general relativity and
has been approached in many different ways. Here we review an
approach taking into account the non-Boolean properties of
quantum logic when forming a spacetime model. Finally, we recall
how the fundamental gauge of diffeomorphisms (the issue of
general covariance vs coordinate conditions) raised deep
conceptual problems for Einstein in his early development of
general relativity. This is clearly illustrated in the notorious
``hole'' argument. This scenario, which does not seem to be
widely known to practicing relativists, is nevertheless still
interesting in terms of its impact for fundamental gauge issues.

\hrulefill
\end{minipage}

\vspace*{1cm}
\section{\large\bf Introduction}
Gauge theories, to which  Friedrich Hehl has contributed so
much, explore the mysterious fundamental role which
symmetries play in our understanding of the physical world.
To have a symmetry we need two parts: something fixed while
something else changes.  Much of the progress of modern
physical theories has come as a result of studying this
``fixed/changing''
dichotomy, analyzing it and suggesting new paradigms.  For
most of the history of physics, space, and more recently
spacetime, has in some sense or other been the underlying
fixed object on which theories are written and in terms of
which experimental results are reported.  The active,
changing, part of the symmetry, the gauge group, consists of
coordinate changes, or to use more contemporary terminology,
diffeomorphisms. Yet, in spite of its uncontrovertible
central role, a thorough understanding of spacetime models
is still one of the most elusive goals of modern physics.
\par In this paper, I would like to review questions related
to these issues using the now discredited ether models of
the eighteenth and nineteenth centuries for comparison. In
the earlier times, the ether was some
not-directly-observable-substratum thought to be needed by
certain theories. For example, in the Newtonian
gravitational precursor to field theory, Newton thought that
action at a distance  was ``... so great an
absurdity , that...no man, who has in philosophical  matters
a competent faculty for thinking, can ever fall into it.''
\cite{Newton}. He speculated that in the case of
gravitation, the force may be produced by varying densities
of the mechanical ether in the presence of gravitating
masses. In later periods it played a more passive role,
providing a fixed, preferred reference system
 relative to which velocities should be
measured for calculations of Lorentz forces and current
sources in Maxwell equations. We will not be concerned with
the actual details of these old ether models here, but only
use them as a backdrop to consider contemporary
questions revolving around observability issues in
physical models.  The interested reader is
invited to consult the massive two volume work on the
ether by Whittaker \cite{Whittaker51}.

\par Those of us who work day by day in
theoretical physics and especially relativity may tend to
take for granted the huge package of assumptions that we
impose on our spacetime models, most of which cannot be
supported by direct experiment.  It is this tacit acceptance
of unobservable properties of our model that motivates this
paper.  Of course, these assumptions have been questioned by
many workers from
the time of Greek physics to the present. Max Jammer  has
given us an excellent review of this subject \cite{Jammer93}.
  Without any claim
to completeness we can also note more contemporary work, for
example, Connes \cite{Connes94}, Rovelli \cite{Rovelli91},
Madore and Saeger \cite{Madore97}, Heller and Sasin
\cite{Heller97}, Brans \cite{Brans80}. Other participants of
this meeting have contributed to this field, including
Rosenbaum \cite{Rosenbaum97} in his lecture to this meeting,
and L\"ammerzahl and Macias \cite{Lamacias94}. \par
Many of these questions border on the philosophical,
and philosophers and historians of science have certainly
made their contributions.  Again the literature is too huge
to survey here.  I mention only the work of Gr\"unbaum
\cite{Gruenbaum77} and Earman \cite{Earman90}.\par
\section{\large\bf Contemporary Spacetime Models}\par In
almost all theories a model of spacetime, say $M$, is
required with at least the following properties:\par
\begin{itemize} \item{\bf Point set}.  That is, $M$ contains
atomic elements, $p\in M$, representing idealized point
events.  The existence and identity of this structure is
absolutely necessary for all of the following ones.
\item{\bf Topological manifold}.  $M$ must have a topology such
that it is locally Euclidean.  That is, each $p$ must lie in a
neighborhood homeomorphic to $\mathbb{R}^n$ for some integer $n$.
Again, this property is needed for the others.
\item{\bf Smooth manifold}. $M$,  in addition to being
locally {\it homeomorphic} to $\mathbb{R}^n,$ must be locally
{\it diffeomorphic} to it.  The local diffeomorphisms
constitute the local coordinates needed to express smooth
functions and to operate on them differentially.
\item{\bf Geometric manifold}.  $M$ must carry a smooth metric,
connection and perhaps, as Hehl has taught us, torsion.
\item{\bf Bundle structures}.  Finally, additional gauge
structures and fields require local patching together of products
of $M$ with models of the field/gauge group space.\end{itemize}
\par
The advent of supersymmetry, etc. surely makes this list
incomplete, but it does provide some idea of the extensive and
detailed structure that spacetime models use. Here we want to
point out that this involved logical construction is built
from bricks that are as essentially unobservable as were
the vortices and atoms of the mechanistic ether of earlier
times. It is in this sense that the title of this talk
compares spacetime to the ether.  But before getting into
the details of these foundational lacunae, let us look at
the basic gauge group of spacetime relativity, starting from
its historical roots.\par
  \section{\large\bf The Classical Ether}\par
Trautman \cite{Trautman70} has suggested a helpful
way to look at the progress of relativity in spacetime
physics using modern terminology. For more  philosophical
and historical details along this path see Jammer
\cite{Jammer93}. Starting with Aristotelian physics we can
consider the basic spacetime model, $M$, as a simple direct
product of space with time, each having some intrinsic,
absolute properties,
$$
M=space\times time=\mathbb{R}^3\times
\mathbb{R}^1. \eqno{(1)}
$$
 In this context, what passed for dynamics was defined by
 ascribing of natural places to specific types of matter,
``earthy'' things to the earth, etc.\par
Skipping centuries of interesting intellectual history, we
arrive at the relativity of Galilei and Newton. In
actuality, as Jammer points out, Newton was strongly
attached to the idea of some absolute nature for space,
something more along the lines of (1).  Nevertheless the
formalism of his mechanics logically leads to a replacement
of (1) with something like
 $$M=\mbox{Bundle}={\left\{\begin{array}{ccl} {\mbox
space=}\mathbb{R}^3 & \longrightarrow & M\\  &   &
\downarrow\vcenter{\rlap {$p$}}\\ & &
\mathbb{R}^1{\mbox=time}\\ \end{array}\right.}\eqno{(2)}$$
Thus, spacetime,
 $M$ is a  {\it bundle} over time, ${\mathbb{R}^1},$
with space as fiber, $\mathbb{R}^3$.  What distinguishes
bundles from ordinary products as in (1) is the absence of
any {\it natural} identification of the fiber over one base
point with that over some other base point. Each fiber,
``space at a given time,'' is isomorphic to $\mathbb{R}^3$,
but in no natural or canonical way.    In classical
mechanics, the fiber group, relativity gauge group $G,$ is
Galilean group. This is essentially the real linear affine
group of dimension three. We can complete this picture to
describe classical mechanics in modern terms by adding a
preferred flat Euclidean metric on the fiber, three-space,
and a preferred linear structure on the base space, time.
Finally, we need a corresponding bundle connection whose
geodesics provide the paths of free particles.\par
Thus, the formalism of Newtonian mechanics is more naturally
associated with a bundle structure as in (2), rather
than  with absolute space structure, (1).
Nevertheless, Newton felt strongly drawn to the latter,
perhaps in part because of his difficulties with
action-at-a-distance, as illustrated in the quotation in
the Introduction.  Thus, to Newton, space needed some
mechanical properties to enable it to transfer force and
energy over distances in field theories.  This
mechanical structure must be associated with some
``substance,'' the ether, which incidentally provides an
absolute rest frame, or the reduction of the bundle (2) to
the trivial product, (1).\par The attractiveness of such an
absolute space model was reinforced with the  advent of the
unified field theory of electromagnetism.  In this theory
velocity appears twice.  First in the Lorentz force law,
$$
{\bf F}=q({\bf E}+{\bf v\times B})\eqno{(3)},
$$
where {\bf v} is the velocity of the charge $q$ being acted
on by the electromagnetic field, {\bf (E,B)}.  The question
left hanging is: ``velocity relative to what reference
frame?''  Secondly, the velocity {\bf v} appears in the
source field equation
$$
\nabla\times{\bf B}= \mu_0\rho{\bf v}+\frac{1}{c^2}
\frac{\partial{\bf E}}{\partial t},\eqno{(4)}
$$
along with the new quantity of dimension speed,
$c\equiv 1/\sqrt{\epsilon_0\mu_0}.$ Again, the question of
what reference frame should be used to measure the source
current velocity arises.  Furthermore the new, unexpected
 speed, $c$, reappears as the speed of
electromagnetic waves in one of the consequences of the
vacuum field equations, $$ (\nabla^2-\frac{1}
{c^2}\frac{\partial^2}{\partial
t^2})\left\{\begin{array}{l}{\bf E}\\{\bf B}
\end{array}\right\}=0.\eqno{(5)} $$
So, Maxwell's unified field theory of electromagnetism
leaves us with three speeds, that of the source of the
fields, (4), that of the object on which the fields act,
(3), and the field waves themselves, (5). The Galilean
relativity in (2) must then break down, since the presence
of such speeds breaks its invariance, and we return to some
absolute space model, (1), where space is now the
``luminiferous'' ether, with spacetime $$
M=ether\times time. \eqno{(6)}
$$
\par
Of course, there were strong voices in opposition to the
notion of absolute space, most notable Bishop Berkeley
\cite{BerkeleyJ}  and
later Mach who referred to the notion of absolute space as a
``conceptual monstrosity'' \cite{MachJ}.  Einstein claimed
that such ideas were instrumental in the evolution of his
thinking about relativity. \par At this point it might be
appropriate to recall all of the effort that was put into
the design of mechanical or pseudo-mechanical models of such
an ether \cite{Whittaker51}.  It is natural to wonder how
all of the work of contemporary physics involving elaborate
spacetime structures and superstructures may likewise appear
in the next century. \par
\section{\large\bf End of the Classical Ether: Special
Relativity}  But of course some hundred years ago
Michelson and Morley results forced serious rethinking of
the classical ether-space model, (6). While Lorentz and
others attempted to preserve the ether by proposing length
contractions and clock dilations as a result of motion
through it, Einstein cut to the heart of the matter in his
principle of special relativity, closely tied to the
principle of operationalism which informally claims that if
something conspires successfully against its observation,
then its existence should not be used as part of a
physical theory.  Thus, if the ether's only claim to
existence is as an absolute rest reference frame, and its
properties make motion through it unobservable, then from
the viewpoint of physics it doesn't exist. \par
In fact,  Einstein taught us to
think in terms of a unified spacetime model, with no
preferred a priori splitting (apart from the qualitative
space-like, time-like, light like ones) of space from time.
The transformation group preserving these spacetime
properties, the gauge group of special relativity is of
course the Poincar\'e group, that is, the homogenous Lorentz
group plus translations. Friedrich Hehl and his
colleagues have been leaders in emphasizing the
importance of this group especially in the context of
general relativity and its generalizations,
\cite{Hehl76},\cite{Mielke87},\cite{Hehl95}.\par With the
many successes of special relativity, it seems that the
ether has finally been put to rest. Indeed it has in this
classical sense. If you
can't observe it, it doesn't exist is a standard motto. Or
to paraphrase an old axiom:``{   No stuff has existence
until it is observed to have existence.}'' But, should we
not apply this to ``stuff''=manifold properties? So, is
spacetime the new ether? Clearly, it does not play the same
mechanical role of ``transmitter of forces,'' as the vortex
constituted stuff of the old mechanical one. Also, it
clearly does not provide an ``absolute rest reference
frame.'' But, it does have other, similar properties. It
provides, in operationally unobservable ways, the substratum
to carry the many structures used by modern theories.
and it is the point of this paper that
spacetime structures in modern theories comprise a
replacement for it and so have become a ``new ether.''
 \par
\section{\large\bf General Covariance: Einstein's Relativity
}\par In addition to the Principle of Equivalence, which we
will not consider, the Principle of General
Relativity and Mach's Principle, are generally taken as
foundations in models of how Einstein arrived at General
Relativity.  Of course, the actual history is more
complicated and interesting, and the reader can consult the
volume one of the Einstein Studies, \cite{Howard89}, for
a deep and accurate account of the story.\par
For our purposes, it is sufficient  to point out that
Einstein was aware of the rigid structure  still remaining
on the spacetime of special relativity by the Lorentz metric
and the associated preferred set of inertial reference
frames. Mach's Principle addresses the issue of why the
fixed stars have constant velocity in the inertial frames,
while the Principle of General Relativity proposes extending
the physically acceptable frames beyond this restricted set.
In other words, while special relativity had weakened the
assumption of a preferred (zero) absolute-velocity-defining
ether, it replaced it by a preferred (zero)
absolute-acceleration-defining one. So, in the
spirit of this paper, the next step
toward generally covariant theories was a result of
re-examining and loosening previous rigid
structures.  John Norton \cite{Norton89} has given
us an thorough and highly interesting analysis of
how Einstein arrived at his equations of General
Relativity.  Here we will only skim over the issue
of the identity of spacetime points as illustrated in
Einstein's hole dilemma. (See also \cite{Rovelli91}).\par
Consider a model universe, with matter and metric fields,
$T,g$ on a manifold containing a region $U,$ which will be
the ``hole.'' Einstein was thinking in coordinates, so let
$T(x),g(x)$ be expression of solution field equations in
terms of coordinates (global) $x$.  \par Now re-coordinate,
$x\rightarrow x',$ with $x=x'$ in $U$, but not everywhere.
Then $T'(x'),g'(x')$ is also a solution, with
$$g'(x')=g'(x)=g(x),\eqno{(7)}$$
 within $U$.  But matter and fields
are different outside of $U$. So, Einstein was deflected
from seeking a generally covariant theory since the
following fact  would seem to be paradoxical: {\it matter
outside of $U$ does not determine the fields inside of $U$
uniquely (or vice versa)}.  At this point we must be
cautious about treating this  as  trivial, since we are so
accustomed to accepting  general covariance as an obvious
desideratum. From the viewpoint of development of the
theory, there is more here than confusion about
coordination.\par In fact, it does seem on first glance that
Einstein and Grossman were confused about the expression of
the {\it same} metric merely displayed in {\it different}
coordinates.  For example
 $$ds^2=dx^2+dy^2,\eqno{(8)}$$
 or
$${ds'}^2=\cosh^2(x'){dx'}^2+{dy'}^2.\eqno{(9)}$$
 Clearly
(8) and (9) represent the same metric, and Einstein was
aware of this.  However changing the notation in (9) results
in
 $${ds'}^2=\cosh^2(x)dx^2+dy^2.\eqno{(10)}$$
 If we then {\it identify} the points of the manifold with
the pair $(x,y)$,  then (8) and (10) are truly
different metrics in some sense, although they are
diffeomorphic (isometric). In fact it is possible to define
{\it point}=``ordered pair of numbers,'' not
``diffeomorphism equivalence class of ordered pair of
numbers in each coordinate system.''\par This discussion
highlights the difference between the active and passive
interpretations of the transformation (diffeomorphism).
Actually, as discussed in detail by Norton \cite{Norton89},
Einstein's fourth presentation of this argument shows that
rather than being confused at the difference between (8) and
(9), he was laying the groundwork for the modern
interpretation of diffeomorphisms as physical gauge
transformations.   \par What may be surprising about this is
that it seems to rob the individual points of their
identity, in the absence of a metric. In other words, if
$P_1,P_2$ are two points in a manifold, some diffeomorphism
maps one into the other.  The geometry and fields around
$P_1$ become those around $P_2,$ in physically equivalent
geometric manifolds, so $P_1$ cannot be distinguished from
$P_2.$ Rovelli, \cite{Rovelli91} discusses this subject in
some detail, distinguishing spacetime models as $M_L$,
``local,'' with a particular smoothness and ``atlas'' as
opposed to $M_N$, ``non-local,'' which is equivalence class
of all $M_L$ under diffeomorphisms (gauge
transformations).\footnote{Corresponding to this,
mathematicians distinguish ``smooth structure'' from
``smooth manifold.'' We will discuss this later.} However,
let us recall that this discussion concerns the mathematical
model which is mapped by {\it some} assumed
``diffeomorphism'' onto an absolute point set, spacetime.
In fact, without some underlying pointset, there can be no
notion of diffeomorphism.\par
Nevertheless, the idea remains that the use of
diffeomorphism as physically unobservable gauge ``wipes
out'' the individual identity of points.  In fact, in
discussing his final generally covariant field equations
Einstein said in a letter to Schlick in 1915, `` thereby
time and space lose the last remnant of physical reality.
All that remains is that the world is to be conceived as a
four-dimensional (hyperbolic) continuum of 4 dimensions.''
\cite{Einstein15}. Our point here is that this
continuum carries at least as much structure as the
replaced ether.\par

 \section{\large\bf  Absolute Spacetime:
Quantum Theory }
There still remains the list  of
 absolute spacetime properties described in the
introduction such as topology,
smoothness, etc., which seem to be arbitrarily chosen. This
leads to the question of regarding the role of  spacetime
as object or scratch pad. This is a question
certainly bordering on philosophy philosophy, but also
closely related to the operational foundations of
quantum theory. \par
The principal distinguishing characteristic of  quantum
logic as opposed to classical logic is that in quantum
theory questions correspond to projection operators in
Hilbert space. The logical operators, ``or,'' $\vee,$
corresponds to ``span of vector space union,'' while
``and,'' $\wedge,$ corresponds to intersection. This
results in a non-Boolean algebra,
 $$a\wedge(b\vee c)\ne (a\wedge b)\vee
(a\wedge c),\eqno{(11)}$$
By contrast
classical questions concern ``set-inclusion,'' so
``or,'' $\vee,$ becomes set union,
$\cup$, and ``and,'' $\wedge,$ becomes $\cap$, set
intersection. Thus, for point sets,
$$a\cap(b\cup
c)=(a\cap b)\cup(a\cap c).\eqno{(12)}$$
In other words, {\it  quantum logic is not in general
consistent with the logic of set-inclusion, which is
fundamental to point-set questions.}
\par
For further discussion
of these questions, see Brans\cite{Brans80}, and especially
Marlow \cite{Marlow80}. Maybe there is not enough in this
bare-bones quantum logic approach to work with (produce a
theory), but, many others,  too many to mention here, have
also looked into the influence of quantum theory on
spacetime point set properties in different ways. However
the contributions  of Connes \cite{Connes94} and Madore
\cite{Madore95} stand out as leading to much current work in
this area.\par \section{\large\bf  Absolute Spacetime:
Choice of Smoothness }  \par Until recently this was thought
to be trivially determined by the topology, at least for
relatively simple manifolds such as $\mathbb{R}^4$. However,
it is not trivial. In fact, some of the most exciting
developments of differential topology recently have come as
a result of what can be termed ``exotic smoothness'' on
spaces of relatively trivial topology, for example
$\mathbb{R}^4.$ In many respects, the development of this
subject parallels that of non-Euclidean and then
differential geometry.  Thus, for many years there were
conjectures about the uniqueness of Euclidean geometry, both
mathematically and as physics. Similarly, but more recently,
there have been conjectures that there is essentially only
one way to do calculus globally on topologically simple
manifolds.  The phrase ``how to do calculus globally''
corresponds to what mathematicians call a differentiable or
smoothness structure.  Physically such a structure is a
global system of reference frame patches covering all of
spacetime smoothly, that is, with smooth ($C^\infty$)
coordinate transformations in their overlaps. The phrase,
``essentially only one'' means only one equivalence class
under diffeomorphisms of the manifold. This difference
between different smoothness structures and non
diffeomorphic ones can be a slippery concept to master, but
is central to an understanding of differential topology. In
a way, it is parallel to that involved in Einstein's hole
argument discussed above.  Just as the choice of different
coordinates may make the metric look different  when the
underlying geometry is actually the same, so will a
recoordination make the differential structure appear to be
different, when in fact it is equivalent (diffeomorphic).
This equivalence inducing class of diffeomorphisms
corresponds to the underlying principle of general
relativity.  \par By direct calculation, it is possible to
show that, up to diffeomorphisms, there is only one
smoothness structure on each $\mathbb{R}^n$ for $n=1,2,3.$
For $n>4$ the same result was obtained later making use of
cobordism techniques.  However, the case $n=4$ remained
an open one.  Because of the topological triviality it was
natural to conjecture that it too would turn out to be
trivial with respect to differential topology.  Thus, it was
a tremendous surprise when as a result of the work of
Donaldson, Freedman and others it was established that
\par
{\bf Theorem; (Donaldson, Freedman, et al.)}{\it There
are an infinity of smooth manifolds of topology
$\mathbb{R}^4,$ none of which are in the diffeomorphism
class of the any other (including the standard
one).} \par Thus, the diffeomorphism gauge does
not cover the entire range of physics on topologically
trivial $\mathbb{R}^4$!  Do these provide new structures for
new physics? See Brans \cite{Brans94} for a general review
of these topics, and Asselmeyer \cite{Asselmeyer97} for
a specific suggestion of physical content.\par

\section{\large\bf  Conclusions }\par
In this paper we tried to survey some of the extensive
structures used on contemporary spacetime models, noting
their direct physical unobservability and reflect on this
rigid fine structure in the light of the historical parallel
of the luminiferous ether.
A hundred or so years ago, it was generally thought (apart
perhaps from a few people like Gauss) that geometry
was ``pre-physics,'' a natural given. Now, we have take it
for granted that geometric structures carry physical fields.
\par
Perhaps it is now appropriate to speculate that the
mathematical structures  ``point set'' (e.g., Boolean, or
non-Boolean...), ``topological'' (e.g.,
non-Euclidean...), ``smoothness'' (e.g., exotic...), etc.,
might also serve to carry physical properties in a manner
analogous to that of ``geometry.''  \par
 \section{\large\bf  Acknowledgments }
I am  indebted to John Norton and Carlo Rovelli for
very helpful comments, suggestions and insights. Also, this
work was partially supported by a grant, LaSpace,
R150253.\par Finally, of course, we are all grateful to
Friedrich Hehl for his persistent clarification of the role
of various spacetime gauge structures.\par

   \end{document}